\documentclass[floats,aps,nofootinbib,preprint]{revtex4}
\usepackage{bm,amsfonts, mathtools}
\usepackage{graphicx,color, wasysym}
\usepackage{textcomp}
\usepackage{amsmath,amssymb,latexsym,epsfig}

\begin{document}

\makeatletter

\title{Generating Functions for Laguerre Polynomials: New Identities for Lacunary Series}

\author{D. Babusci}
\email{danilo.babusci@lnf.infn.it}
\affiliation{INFN - Laboratori Nazionali di Frascati, via E. Fermi, 40, IT 00044 Frascati (Roma), Italy}

\author{G. Dattoli}
\email{dattoli@frascati.enea.it}
\affiliation{ENEA - Centro Ricerche Frascati, via E. Fermi, 45, IT 00044 Frascati (Roma), Italy}

\author{K.~G\'{o}rska}
\email{kasia_gorska@o2.pl}
\affiliation{H. Niewodnicza\'{n}ski Institute of Nuclear Physics, Polish Academy of Sciences, ul.Eljasza-Radzikowskiego 152, 
PL 31342 Krak\'{o}w, Poland}

\author{K.~A.~Penson}
\email{penson@lptl.jussieu.fr}
\affiliation{Laboratoire de Physique Th\'eorique de la Mati\`{e}re Condens\'{e}e,\\
Universit\'e Pierre et Marie Curie, CNRS UMR 7600\\
Tour 13 - 5i\`{e}me \'et., B.C. 121, 4 pl. Jussieu, F 75252 Paris Cedex 05, France\vspace{2mm}}

\begin{abstract}
We present a number of identities involving standard and associated Laguerre polynomials. They include double-, and triple-lacunary, ordinary and exponential generating functions of certain classes of Laguerre polynomials.
\end{abstract}

%--------------------------------------
\maketitle
%-----------------------------------------------------------

The purpose of this note is to list a number of identities satisfied by standard Laguerre polynomials $L_{n}(x)$ and their associated counterparts $L_{n}^{(\alpha)}(x)$, with $L_{n}^{(0)}(x) = L_{n}(x)$. For $L_{n}^{(\alpha)}(x)$ as well as for the Hermite polynomials $H_{n}(x)$ we employ the definitions given in \cite{1}. We present these identities below:
\begin{eqnarray}\label{eq1}
\sum_{n=0}^{\infty} \frac{t^{n}}{n!} L_{2 n}(x) &=& {\rm e}^{t} \sum_{r=0}^{\infty} \frac{(i x \sqrt{t})^{r}}{(r!)^{2}} H_{r}(i\sqrt{t}); \\[0.9\baselineskip]
\sum_{n=0}^{\infty} \frac{t^{n}}{n!} L_{2n}^{(1)}(x) &=& {\rm e}^{t}\sum_{r=0}^{\infty} \frac{p_{2}(r; x, t)}{r! (r+3)!} (i x\sqrt{t})^{r} H_{r}(i \sqrt{t}),  \label{eq2a} \\[0.6\baselineskip]
p_{2}(r; x, t) &=& (1+2 t)r^{2} + (5-4x t + 10 t) r + (6 + 12t - 12 x t + 2 t x^{2}); \label{eq2b} \\[0.9\baselineskip]
\sum_{n=0}^{\infty} \frac{t^{n}}{n!} L_{2n}^{(2)}(x) &=& {\rm e}^{t}\sum_{r=0}^{\infty} \frac{p_{4}(r; x, t)}{r! (r+6)!} (i x\sqrt{t})^{r} H_{r}(i \sqrt{t}), \label{eq3a} \\[0.6\baselineskip]
p_{4}(r; x, t) &=& (2+10t+4t^{2})r^{4} + [36 + (180 - 20 x) t  + (72 - 16 x) t^{2}]r^{3} \nonumber \\
&+& [238 + (1190 - 300 x + 10x^{2})t + (476 - 240 x + 24 x^{2})t^{2}] r^{2} \nonumber \\
&+& [684 +(3420 - 1480 x + 110x^{2}) t + (1368 - 1184 x + 264 x^{2} - 16 x^{3})t^{2}] r \nonumber \\
&+& 720+(3600-2400x+300x^{2})t \nonumber \\
&+& (1440 - 1920 x + 720 x^{2} - 96 x^{3} + 4 x^{4})t^{2}; \label{eq3b} \\[0.9\baselineskip]
\sum_{n=0}^{\infty} \frac{t^{n}}{n!} L_{2 n + k}(x) &=& {\rm e}^{t} k! \sum_{r=0}^{\infty} \frac{(i x \sqrt{t})^{r}}{r! (r+k)!} L_{k}^{(r)}(x) H_{r}(i \sqrt{t}); \label{eq4} \\[0.9\baselineskip]
\sum_{n=0}^{\infty} \frac{t^n}{n!} L_{3 n + k}(x) &=& {\rm e}^t k! \sum_{r=0}^{\infty} \frac{L_{k}^{(r)}(x)}{(r+k)!} \left[\sum_{s=0}^{[r/3]} \frac{(- t x^3)^s (i x \sqrt{3 t})^{r - 3 s}}{s! (r - 3 s)!} H_{r-3s}\left(i \frac{\sqrt{3t}}{2}\right)\right]; \label{eq5} \\[0.9\baselineskip]
\sum_{n=0}^{\infty} \frac{t^n}{n!} L_{3 n}^{(1)}(x) &=& {\rm e}^t \sum_{r=0}^{\infty} \frac{q_{3}(r; x, t)}{(r+4)!} \left[\sum_{s=0}^{[r/3]} \frac{(- t x^3)^s (i x \sqrt{3 t})^{r - 3 s}}{s! (r - 3 s)!} H_{r-3s}\left(i \frac{\sqrt{3t}}{2}\right)\right]; \label{eq6a} \\[0.6\baselineskip]
q_{3}(r; x, t) &=& (1+3t) r^3 + (9 + 27t - 9tx)r^2 + (26 + 78t -63tx + 9 t x^2)r \nonumber\\
&+& (24 + 72t -108tx +36tx^2-3tx^3); \label{eq6}
\end{eqnarray}
In Eqs.~\eqref{eq5} and \eqref{eq6a} $[n]$ is the floor function.
\begin{eqnarray}
\sum_{n=0}^{\infty} \frac{\left(\frac{1}{2}\right)_{n} t^n}{\left(1+\frac{\alpha}{2}\right)_n} L_{2n}^{(\alpha)}(x) &=& (1-t)^{-\frac{1+\alpha}{2}} \sum_{r=0}^{\infty} \frac{L_{r}^{(r+\alpha)}\left(\frac{x}{2}\right)}{\left(1+\frac{\alpha}{2}\right)_r} \left[-\frac{tx}{2(1-t)}\right]^r; \label{eq7} \\[0.9\baselineskip]
\sum_{n=0}^{\infty} \frac{\left(\frac{1}{2}\right)_{n}}{(1+m)_{n}} t^n L_{2 n}^{(2 m)}(x) &=& \frac{1}{\sqrt{1-t}} \left(\frac{x 
\sqrt{t}}{2}\right)^{-m} \exp\left(-\frac{t x}{1-t}\right) I_{m}\left(\frac{x \sqrt{t}}{1-t}\right), \label{eq8} 
\end{eqnarray}
where $m=0, 1, 2, \ldots$ and $I_{m}(z)$ is the modified Bessel function;
\begin{eqnarray}
\sum_{n=0}^{\infty} \frac{\left(\frac{1}{3}\right)_{n} \left(\frac{2}{3}\right)_{n} t^n L_{3 n}^{(\alpha)}(x)}{\left(1+\frac{\alpha}{3}\right)_{n} \left(\frac{2}{3}+\frac{\alpha}{3}\right)_{n}} &=& (1-t)^{-\frac{1+\alpha}{3}} \sum_{r=0}^{\infty} \frac{\Gamma(3r+\alpha+1)}{\left(1+\frac{\alpha}{3}\right)_{r} \left(\frac{2}{3}+\frac{\alpha}{3}\right)_{r}} \left[-\frac{tx}{9(1-t)}\right]^r \nonumber \\[0.3\baselineskip]
&\times& \left[\sum_{s=0}^{r} \frac{(-x)^s L_{s}^{(s+\alpha+r)}\left(\frac{x}{3}\right)}{(r-s)!\; \Gamma(2s + \alpha+r+1)}\right]; \label{eq9} \\[0.9\baselineskip]
\sum_{n=0}^{\infty} t^n L_{2 n}(x) &=& \frac{1}{1-t} \sum_{r=0}^{\infty} \frac{L_{r}^{(r)}\left(\frac{x}{2}\right)}{\left(\frac{1}{2}\right)_r} \left[-\frac{tx}{2(1-t)}\right]^r; \label{eq10} \\[0.9\baselineskip]
\sum_{n=0}^{\infty} t^n L_{3 n}(x) &=& \frac{1}{1-t} \sum_{r=0}^{\infty} \left(-\frac{3 t x}{1-t}\right)^r \left[\sum_{s=0}^{r} \frac{r! (-x)^s}{(r-s)! (r + 2s)!} L_{s}^{(s+r)}\left(\frac{x}{3}\right)\right]; \qquad \label{eq11} \\[0.9\baselineskip]
\sum_{n=0}^{\infty} t^n L_{2n}^{(\alpha-2n)}(x) &=& (1-t)^{\frac{\alpha}{2}} \cosh\left[\sqrt{t}x - i \alpha\arcsin\left(\frac{\sqrt{t}}{\sqrt{t-1}}\right)\right]. \label{eq12} 
\end{eqnarray}
To the best of our knowledge all the above formulas are new. The Eq.~\eqref{eq8} is the corrected version of Eq.~(5.11.2.10), p. 704 of \cite{1}. We shall present the detailed derivation of identities Eqs. \eqref{eq1}-\eqref{eq12} in the forthcoming publication \cite{2}.

We thank A. Bostan, D. Foata and V. Strehl for illuminating discussions. K.~G. and K.~A.~P. acknowledge support from Agence Nationale de la Recherche (Paris, France) under Program PHYSCOMB No. ANR-08-BLAN-0243-2.

\end{document}